# Military and terrorist attacks against chemical weapons sites and the prospect of a "Syrian War Syndrome"

Theodore E. Liolios[1], Konstantinos G. Kolovos[2]

**Abstract:** Military and terrorist attacks or accidental explosions on chemical weapons storage sites such as research centers, ammunition depots and factories can explosively release large quantities of lethal chemical agents which can affect not only the vicinity of ground zero but also inhabited and cultivated areas at large distances. The simulations carried out in this work focus on the Sarin warfare agent while the explosive agent releases are assumed to occur at night (usually preferred by the attackers for obvious reasons). The exposed population downwind is assumed to be unprotected, in an open area (e.g. in the streets), without any warning and thus receptors can remain immersed in the Sarin cloud for considerable time (at least ten minutes) before any protective action is taken (evacuation, finding shelter, receiving medical care etc.). The results indicate that the effects of military, terrorist and accidental explosions on Sarin storage areas could be devastating at large distances from ground zero as they would practically amount to gigantic lethal chemical weapon explosions. The models of this work are implemented on a case study, namely the April 14, 2018 military strikes on the alleged Syrian chemical weapons sites due to its high relevance and similarity to the Sarin releases occurred in the US demolition operations at the Khamisiyah Pit in Iraq (1991) believed to have been a possible source of the "Gulf War Syndrome". According to this case study if Sarin agent was indeed stored in the alleged Syrian chemical weapons sites then various populated areas around ground zero would have experienced lethal or life-threatening effects, irreversible or other serious long-lasting adverse health effects or at least notable discomfort. Moreover, if the Khamisiyah Pit Sarin ammunitions demolition operation was indeed the source of the "Gulf War Syndrome" then the incontrovertible multiple uses of Sarin against the Syrian population (possibly aggravated by the April 14, 2018 attacks) might give rise to a similar "Syrian War Syndrome" which is likely to appear in the future. Forensically, if after the bombardment of the alleged Sarin storage sites in Syria there are no symptoms of exposure to Sarin in populated areas close to ground zero (predicted and mapped by the postulated scenaria of the present work) then that may be a strong indication that the attacks were probably unjustified and unfair.

[1] *(a) Professor and Director of the Physical Sciences Department at the Hellenic Military Academy (http://sse.army.gr/en/), Director of the Arms Control Center (www.ArmsControl.eu).*
*(Corresponding author, phone: +30 6944165341, e-mail: Director@ArmsControl.gr)*

[2] *(b) Assistant Professor of the Physical Sciences Department at the Hellenic Military Academy (http://sse.army.gr/en/), Senior Researcher at the Arms Control Center (www.ArmsControl.eu).*
*(phone: +30 6972133503,*
*e-mail: kolovosk@gmail.com)*

***Keywords:*** *Chemical Weapon, Chemical Warfare, Nerve Agents, Chemical Warfare Agents, Sarin, Chlorine, EPIcode, Syrian War, Gulf War Syndrome, Syrian War Syndrome, Arms Control*

I. INTRODUCTION

There is incontrovertible proof that chemical weapons, especially Sarin and Chlorine, have been recently used in Syria against innocent civilians. The civilized world has been alerted to these atrocities and has justifiably threatened to punish those responsible for committing crimes against humanity. This work will not be concerned with the identification of the perpetrators who have not been identi-fied beyond reasonable doubt. Responding to that murderous employment of weapons of mass destruction the USA, the UK and France have launched military strikes against Syrian targets reported as chemical weapons sites. However, military and terrorist attacks against such targets or even accidental explosions and demolitions of chemical weapons can have catastrophic effects and produce mass casualties and environmental disaster at large distances from ground zero. It is the purpose of this work to study such events, focusing on the attacks against the alleged targets in Syria.

The critical difference between the effects of combat use of chemical munitions and a bombardment or a terrorist attack on a chemical weapons site are the large quantities of chemical agents and explosives involved in the latter case. In a limited chemical weapon attack single artillery shells, bombs or missiles are involved where the chemical agent is dispersed by means of relatively low energy explosions (from a few hundred grams to a few kilograms of TNT equivalent). When a chemical factory or an extensive chemical weapons depot is attacked then the source term is orders of magnitude larger than the quantities filling chemical weapons munitions. Moreover, conventionally armed missiles, bombs or VBIEDs can carry hundreds of kilograms of TNT equivalent and when detonated on a chemical weapons depot or factory they may also cause the sympathetic detonation of the high explosives inside the chemical munitions stored in the area, thus creating a powerful mechanism of initial chemical agent dispersion. It has been argued before (Research Advisory Committee on Gulf War Veterans' Illnesses, 2008) that the so-called Gulf War Syndrome was not caused exclusively by the Depleted Uranium used by the Allied Forces against Iraq but it may have resulted from large scale chemical agent releases in the atmosphere following the attacks on Iraq's chemical weapons facilities such as research centers, factories and depots. Paramount among the possible sources of the Gulf War Syndrome is the well-known "Demolition Operation at the Khamisiyah Pit". The US DOD has published (US DOD Technical Report, 2002) an extensive report of the event where an attempt has been made to model the explosive release of Sarin and Cyclosarin at Khamisiyah and the subsequent harmful effects on US troops. The preceding arguments suffice to warrant an estimation of the effects of a military or a terrorist attack on chemical weapons sites in order to inform the international community and especially policy and decisions makers about the consequences. As Sarin effects are orders of magnitude more lethal than those of Chlorine the latter will be investigated in a subsequent study.

## II. MODELLING AND SIMULATION

The attacks will be simulated by using the EPICODE 8.0.2 (Homann & Alluzi, 2016) software which is extensively used by the Arms Control Center and is also currently a standard teaching tool in the author's lectures at the Hellenic Army Academy (Weapons Sciences Course). As there is no credible information about the existence or the quantities of chemical agents stored in the sites attacked the source term of the Sarin releases modelled in this work will be selected according to the chemical agent quantities stored in an average chemical weapons facility.

### A. Source Term and Explosion Energy

Chemical weapons munitions are arranged primarily in the following categories: grenades, mortars, rockets, artillery shells, aerial bombs, spray tanks, chemical mines and others. With the exception of spray tanks all other means of delivery use high explosives (fuse+burster) to disperse chemical agents on targets. The total quantity of the chemical agent filling each munition and the high explosives used to disperse the agents varies from fractions of a kilogram to hundreds of kilograms (US Army Chemical Materiel Destruction Agency, 1994). On the other hand, the total quantity of high explosives in the explosive train of the munitions (fuse+burster) used to disperse the agents usually varies from fractions of a kilogram to roughly ten kilograms (US Army Chemical Materiel Destruction Agency, 1994; TM 9-1325-200, 1966). Chemical weapons storage and disposal sites around the world may store hundreds or thousands of tons of chemical agents either in bulk containers or as chemical weapons munitions (National Reseach Council, 1984). Regarding the weapons of the attacker there is a wide range of possibilities. The explosion energy of modern artillery shells, missiles, aerial bombs, IEDs and VBIEDs ranges from a few kilograms to hundreds of kilograms of TNT equivalent (MOABs not included). For example a cruise Tomahawk missile (US NAVY, 2018) reportedly carries a 1000 lb high explosive warhead (unitary warhead with penetrator), a typical GP 2000 lb aerial bomb carries (TM 43-0001-28, 1996; Hyde, 1997) approximately half a ton of TNT equivalent while artillery shells carry up to a few tens of kilograms of TNT equivalent (e.g. US M106 8-inch shells filled with 38.8 lb of Comp B) (Hyde, 1997; TM 43-0001-28, 1996). IEDs and VBIED's explosion energies range from a few kilograms of TNT equivalent (pipe bombs, suitcase bombs, backpacks etc) to thousands of kilograms of TNT equivalent (e.g. trucks loaded with plastic explosives, ANFO or LPG) (FEMA, 2011). Note that modern weapons precise technical data are classified but this is immaterial to the present study as they do not differ significantly from their older variants in terms of explosion energy and chemical fillings.

The nature of the chemical agents themselves associated with the attack or the accident play a significant role in the ensuing harmful effects. Non-persistent agents such as Sarin (GB) will quickly evaporate (especially in warm weather areas) and will not pose a risk to humans after the chemical cloud and the nearby pools have evaporated and the vapors have been diluted into higher layers of the atmosphere. Other persistent agents, however, such as Mustard Gas (HD) or VX will contaminate the area for a long time denying entrance to unprotected personnel and requiring large-scale decontamination efforts. Chlorine dispersions, on the other hand, pose a much less significant risk than GB, HD, and VX and will not prevent first responders from entering the contaminated areas if they are equipped with protective gear. This work focuses on one of the most lethal agents used in Iraq and Syria, that is Sarin (GB), and it is only the beginning of a series of reports regarding the effects of other important chemical weapons agents used in combat or against innocent civilians.

All the scenaria in this work assume that a chemical weapons site storing significant quantities of Sarin (GB) is attacked (or suffers an accident) resulting in a chemical explosion that completely (100% airborne) disperses large quantities of Sarin. The maximum concentration downwind from ground zero is proportional to the initial source term dispersed in the explosion (according to the Gaussian dispersion model used by EPICODE). Thus, if larger or smaller quantities are dispersed in the attack the derived downwind concentrations of the present simulations may be simply multiplied by an appropriate multiplicative factor to obtain a new estimation (larger or smaller source terms). Plausible TNT equivalents are selected in all cases according to the above analysis. Nevertheless, we can always bracket the effects by increasing/decreasing the TNT equivalent by an order of magnitude (e.g. 10kg TNT to 100kg TNT) and observe the sensitivity of the effects to such a variation.

### B. Meteorological Parameters

All simulations in this study assume very unfavorable weather conditions which considerably increase the concentration of the chemical agents close to the ground, thus a few comments regarding unfavorable and worst-case atmospheric scenaria are necessary. Worst-case scenaria set upper limits of possible preventive and mitigation procedures while average reasonably unfavorable scenaria are indicative of the risk associated with the problem at hand. For example finding the absolutely worst-case input parameters (an extremely difficult task) may lead to unrealistic modelling and suggest exaggerated preventive and mitigations actions. Simple, plausible, unfavorable scenaria, however, can sometimes adequately illustrate the magnitude of the problem without risking being labeled extreme and unrealistic.

The Gaussian dispersion model used by EPICODE for point sources on the ground yields maximum downwind concentrations for low wind speeds (concentrations are inversely proportional to wind speed) and moderately stable atmospheres (e.g. stability class F) for a wide range of deposition velocities. However, extremely small wind speed values will yield unrealistic results especially for zero deposition velocities. Even the US Environmental Protection Agency (EPA) recommends (EPA-454/R-99-005, 2000) the use of a minimum wind speed threshold of 0.5 m/s. In addition, explosives releases in EPICODE (Homann & Alluzi, 2016) and HOTSPOT (Homann & Aluzzi, 2013) do not yield downwind concentrations which are inversely proportional to the wind speed very close to ground zero due to the complex form of the source term which is no longer point-like but an extended source whose geometry is determined by the explosion energy.

Consequently, it should be emphasized that predictive Gaussian model simulations in arms control studies must be a combination of realism and pessimism (unfavorable scenario adoption), while reconstructive simulations should input the most reliable available input parameters from credible sources. For example, for obvious reasons, the attacker in most cases would prefer to launch his attack at night which enhances the probability of F, E and D Pasquill stability classes. For many practical purposes a Gaussian plume model with a stability class F and a low wind speed (such as one meter per second) yields sufficiently conservative results for ground releases (Homann & Aluzzi, 2013).

### C. Deposition Velocity

When particulate and gaseous materials are dispersed in the atmosphere, they are transferred to the ground surface through a variety of mechanisms collectively described as dry deposition (in the absence of precipitation) (Sugiyama, et al., 2014). Accordingly, Sarin, after its explosive release (in the form of aerosols or vapors), will be deposited on the ground (and other surfaces such as trees, water etc.) due to gravitational settling, turbulent diffusion and Brownian motion while various other chemical and biological processes will contribute to the plume depletion. A crucial parameter in the dispersion of Sarin (and in the dispersion of any other atmospheric pollutant) is its deposition velocity, which for most gaseous and particulate materials dispersed into the atmosphere vary from $v_d$=0.001 cm/sec to $v_d$=10 cm/sec (Sugiyama, et al., 2014).

Sarin has a boiling point slightly larger than that of water, and when exposed to the high temperatures of an explosion (thousand degrees Celsius) a large quantity of the agent will be vaporized and dispersed with a practically zero deposition velocity. The rest of the agent will be either aerosolized and dispersed with small deposition velocities (similar to those associated with an explosive release of hot water aerosols) or form pools and hotspots in the vicinity of ground zero and continue to evaporate after the attack. Note that we are only concerned with dry deposition disregarding for the time being wet deposition velocities (i.e. due to precipitation) and any other mechanism of agent removal from the atmosphere. Large deposition velocities increase the concentrations close to ground zero and deplete the cloud at large distances while very small deposition velocities (e.g. associated with vapors) allow the cloud to reach larger distances. The most conservative assumption (unrealistic) is that the high temperatures of the explosion completely vaporize the entire quantity of Sarin which is then deposited downwind at a deposition velocity close to zero. Such an assumption, however, would lead to unrealistically large concentration estimates downwind even at very large distances.

Hence, it is reasonable to bracket the effects of Sarin dry deposition (aerosol and vapors) by studying the concentrations downwind in particularly unfavorable meteorological conditions assuming deposition velocities from $v_d$=0.0 cm/s to $v_d$=10 cm/sec although the most plausible dry deposition velocity is $v_d$=0.3 cm/s recommended for the respirable component of the source term (particles with less than 10μm aerodynamic diameter) by EPICODE 8.0 (Homann & Alluzi, 2016) ), HOTSPOT 3.0.3 (Homann, 2015) and by the US DoD NARAC software RASCAL (US NRC RASCAL 3.0, 2000). It should be underlined that, even the US DoD modelers (running the codes HPAC/SCIPUFF) for explosive releases of Sarin also assumed a dry gaseous deposition value of 0.3 cm/sec (2000 modelling, (US DOD Technical Report, 2002)). In particular, DHS and CIA field tests (carried out in May 1997) associated with the Khamisiyah incident (US DOD Technical Report, 2002)) suggested that when high explosives detonate close to 122mm rockets filled with Sarin they can cause the sympathetic detonation of the rocket's central burster which will rupture the warhead releasing the liquid chemical agent in the vicinity of the explosion. The bulk of Sarin will form liquid pools soaking into the sand and/or raining onto other surrounding material before slowly evaporating while only a small amount will be instantaneously turned into a cloud of Sarin vapors and aerosol, which will pose a great inhalation hazard due to the high volatility and vapor pressure of Sarin. A typical 122 mm Sarin rocket contained 6.3 kg of chemical agent with an agent purity of approximately 50% and the ration of Sarin to cyclosarin was 3:1. Due to its non-specific nature the present work will assume 100% purity of Sarin (no cyclosarin, no precursors to be mixed before the explosion), with no degradation of the chemical agent although any percentage of impurity or degradation will simply scale the source term and the concentrations downwind.

### D. Explosive dispersion of 100 tons and 100 kg of Sarin with 100 kg of TNT

Let us assume that 100 tons of pure Sarin stored at a large chemical weapons site (National Reseach Council, 1984) is explosively dispersed with 100 kg of TNT. Such a dispersion can be caused by the combined explosion of a Tomahawk missile warhead and the bursters of several chemical weapons munitions. Similar TNT equivalent energy yields can be achieved in terrorist attacks with a vehicle

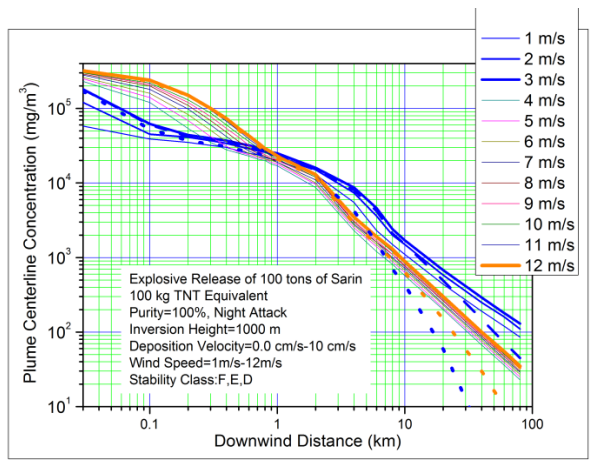

*Figure II.1: Sarin plume centerline concentration with respect to downwind distance from ground zero after 100 tons of Sarin are explosively dispersed with 100 kg TNT equivalent. A night attack is assumed with wind speeds ranging from 1 m/sec to 12 m/sec and a mixing height of 1000 m. Blue lines indicate stability class F, orange lines indicate stability class D. Solid lines indicate deposition velocity $v_d=0$, dashed lines indicate $v_d=0.3$ cm/sec and dotted lines indicate $v_d=10$ cm/sec.*

born improvised explosive device (VBID) such as a car loaded with explosives (FEMA, 2011).

Running a few explosion model simulations with various time-averaged wind speeds ranging from 1 m/s to 12 m/s and focusing only on night-time attacks (stability classes F, E, D and zero deposition velocity $v_d=0$) shows that up to a distance of one kilometer the maximum wind-speed associated with an F class stability (i.e. u=3 m/s, thickest dark blue line) gives lower concentrations downwind than a reasonably high wind speed associated with a D class stability (u=12 m/s, thickest orange line). This trend is reversed for the aforementioned weather parameter combinations at distances larger than one kilometer from ground zero. Although a quantity of 100 tons of Sarin has been used in Figure II.1, the same trends will be observed for any other arbitrary quantity since the concentration of the chemical agent downwind is proportional to the source term (except for ground zero). For example if the quantity of Sarin dispersed in the detonation is one thousand times smaller (i.e. 100 kg) then all quantities on the vertical axis should be divided by 1,000 (see Figure II.2).

Therefore in night attacks there is a competition of worseness (i.e. lethality) between (Stability Class F, u=3 m/s) and (Stability Class D, u>3 m/s). Adopting a more plausible deposition velocity $v_d=0.3$ cm/sec (thick dashed lines) and a reasonable upper limit $v_d=10$ cm/sec (thick dotted lines) we can bracket the possible concentration values downwind by plotting in the same color the curves corresponding to the same meteorological conditions, i.e. three orange lines (solid, dashed, dotted) for (Stability Class D, u=12 m/s) and three blue lines (solid, dashed, dotted) for (Stability Class F, u=3 m/s).

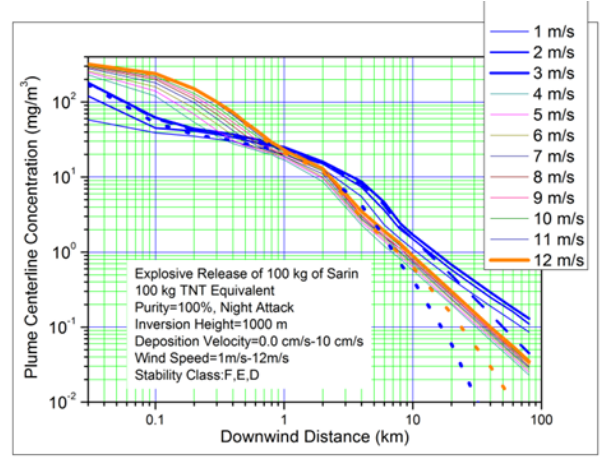

*Figure II.2: Sarin plume centerline concentration with respect to downwind distance from ground zero after 100 kg of Sarin are explosively dispersed with 100 kg TNT equivalent. A night attack is assumed with wind speeds ranging from 1 m/sec to 12 m/sec and a mixing height of 1000 m. Blue lines indicate stability class F, orange lines indicate stability class D. Solid lines indicate deposition velocity $v_d=0$, dashed lines indicate $v_d=0.3$ cm/sec and dotted lines indicate $v_d=10$ cm/sec*

Note that the thick dashed orange line $v_d=0.3$ cm/sec is indistinguishable from its solid orange counterpart ($v_d=0.0$ cm/sec) in Figure II.1 and Figure II.2. On the other hand, the thick blue dotted line in Figure II.1 and Figure II.2 suggests that in the event of night attacks where a quantity of 100 tons (100 kg) of Sarin is dispersed with an explosion of 100 kg TNT there is a probability that individuals up to a distance of 85 km (10 km) downwind from ground zero who remain for ten minutes inside the Sarin plume could experience life-threatening health effects or death as the concentration at their position can possibly be 0.37 mg/m$^3$ (the receptor will be in an AEGL-3 zone).

Assuming a linear risk model the lethal probabilities downwind scale with the quantities of Sarin released therefore the risk at a certain distance from ground zero associated with the dispersion of 100 tons of Sarin is one thousand times larger than the risk associated with the dispersion of 100 kg of Sarin. Regarding the more plausible scenario of the explosive release of 100 kg of Sarin with 100 kg of TNT (Stability Class F, u=3 m/sec, Mixing Height=1000, $v_d=0.3$ cm/sec, see thick dashed blue lines in Figure II.2) EPICODE yields the following output: the AEGL-3 zone (life-threatening health effects or death) can reach distances of 24 km downwind covering an area of 8.9 km$^2$, the AEGL-2 zone (irreversible or other serious, long-lasting adverse health effects) can reach distances of 54 km downwind covering an area of 34 km$^2$ and finally the AEGL-1 zone (discomfort, irritation, or certain asymptomatic non-sensory effects) can reach distances larger than 200 km downwind covering an area of 303 km$^2$. In the event that 100 tons of Sarin are dispersed with the same quantity of TNT EPICODE predicts that all three AEGL zones will exceed a distance of 200 km covering areas as follows AEGL-3 (4000 km$^2$), AEGL-2 (6000 km$^2$), AEGL-1 (8000 km$^2$). The average population density in Syria is approximately 100 people per km$^2$ (2018) therefore the explosive release of 100 kg of Sarin after an attack on its alleged remaining chemical weapons sites could theoretically cause several thousand deaths and tens of thousands of victims suffering irreversible health effects. It is now obvious that a more focused study should be carried out regarding Syria focusing on the alleged chemical weapons sites that were targeted by the allied forces.

III. THE APRIL 14, 2018 MILITARY ATTACKS AGAINST THE ALLEGED CHEMICAL WEAPONS SITES IN SYRIA

The world has witnessed many times the use of chemical weapons (especially Sarin and Chlorine) in Syria against innocent civilians. On April 14, 2018 at 04:00 Syrian time, the United States, the United Kingdom and France launched coordinated missile attacks (mainly with SLCM, ALCM) against alleged chemical weapons sites in Syria. Reportedly, the attacks were in response to the alleged Syrian government use of chemical weapons against innocent civilians - an accusation denied by the Syrian government. The strikes began at 9 pm EDT, April 13 (04:00, April 14, in Syria) targeting with missiles three sites (Rocha, et al., 2018; US DoD, 2018): The Barzah research center located in Barzeh Damascus, an alleged chemical weapons storage facility near Homs (Him Shinshar), and an alleged equipment storage facility and command post also near Homs. The weapon of choice was again (Gearan & Ryan, 2018) the Tomahawk missile whose actual yield (in TNT equivalent) is a significant input parameter in this report. Various other missiles were used during the April 14, 2018 operation such as the joint air-to-surface standoff missiles, Storm Shadow missiles, SCALP cruise missiles etc. (US DoD, 2018). However, their TNT equivalent will be covered by the range of 10 kg to 100 kg TNT that will be postulated for the Tomahawk missiles.

*A. Tomahawk Cruise Missiles Fundamentals*

The main weapon used by the USA in the April 14, 2018 attacks against Syria was the Tomahawk cruise missile (Gearan & Ryan, 2018) which can carry unitary or submunition warheads. Unitary warheads are actually kinetic energy penetrators (extremely hard penetrating devices) carrying as protective capsules inside them a quantity of high explosives to be detonated inside the target. Unitary (kinetic energy) warheads are used against hard and/or buried targets such as weapons production and storage facilities which are relevant to the present study. On the other hand, submunition warheads are actually clusters of bomblets which are used against soft targets such as military air bases with parked aircrafts, trucks, personnel in the open, radar sites, tent cities etc. There are many variants of Tomahawk missiles as of 2014 (BLOCK I,II,III,IV) (Jones, 2014) with various degrees of complexity and sophistication. Unitary warheads, which are of particular interest here, usually have (Hewish, 1998; US NAVY , 2018) an approximate weight of 450 kg, a diameter of about 50 cm and an impact velocity of about 260 m/sec to 335 m/sec. Impact angles vary from 30° to 90° (relative to the target surface).

Tomahawk missiles (Tsipis, 1983; Hewish, 1998; Lewis & Postol, 1992; US NAVY , 2018) penetrate hard targets either relying solely on the kinetic energy of their hardened warhead (kinetic energy penetrator plus high explosives inside) or by using a multiple warhead system which consist of forward-mounted shaped-charge devices and the main kinetic energy penetrator. The precursor devices are designed to precondition the hard target for defeat by either thickening its wall or, if possible, forming a channel along the line of sight so that the follow-through kinetic energy penetrator, which is the main warhead carrying the bulk of high explosives, can penetrate more effectively and detonate inside the target. The precise TNT equivalent of the high explosives carried either by the forward-mounted shaped-charge warheads or by the main kinetic energy penetrator warhead is classified. However, various sources converge on an approximate total weight of 450 kg (Tsipis, 1983; Hewish, 1998; Lewis & Postol, 1992) for the warhead which carries a quantity of (Hewish, 1998) 23 kg to 135 kg of high explosives. Unburnt fuel remaining in the missile at the time of warhead detonation is another crucial parameter as it will be ignited and increase the energy of the explosion (for example a BGM-109 missile carried 272 kg of fuel and modern Tomahawk variants are expected to carry similar amounts). Based on the preceding analysis it is very reasonable to assume that a Tomahawk missile unitary warhead can have a yield ranging at least from 10 kg TNT to 100 kg TNT (the upper limit of this conservative assumption has already been used Section II).

## B. Simulating the missile attacks against the Barzah Research Center

The Barzah research center and the Him Shinshar site near Homs were reportedly attacked on the assumption (US DoD, 2018) that they were chemical weapons sites, with the former allegedly being a chemical weapons research center and the latter a chemical weapons depot. If there was indeed Sarin stored in these sites the Tomahawk kinetic energy penetrators, which totally destroyed the sites according to post-attack footage and public information (US DoD, 2018), would have certainly breached the Sarin containers or the chemical munitions. The high explosives of the warheads (plus the yield of their shaped-charge precursors and the remaining fuel) would have dispersed the chemical agent around ground zero while the chemical cloud generated by the explosion would have been carried downwind. Non-vaporized liquid Sarin would have formed pools and hotspots in the area which would then begin to vaporize creating a residual Sarin plume. Historical weather data for Damascus indicate that (Time & Date, n.d.; US DoD, 2018) at the time of attacks (04:00, April 14, 2018 – Syrian time) in Damascus (approximately 5.5 km south-west of the Barzah center) there was a SW wind with an average speed of approximately 10 km/h (2.77 m/sec) and an average temperature of 12°C (these relatively low temperatures lasted until 07:00 and doubled by 16:00). Given the overall uncertainties of the historical weather data the postulated attack scenario on the Barzah research center will be modelled by assuming the explosive dispersal of one, ten and one hundred kilograms of Sarin with explosions of ten and one hundred kg TNT with deposition velocities ranging from 0.3 cm/sec to 10 cm/sec. According to Section II night hours with low wind speeds (~3 m/sec) suggest a stability class F. We will also assume an average inversion height of 1000 m. The relatively low temperature will lower the plume evaporation rate for hours which is an additional negative factor. In an actual missile attack its warhead will explode after penetrating the structures where Sarin has been stored and therefore the actual release of the agent into the atmosphere is not easily modelled and predictable as has already been underlined. Various factors such as the walls of the target and the ensuing debris from the demolished target would reduce the amount of Sarin which would be rendered airborne. Due to these inevitable uncertainties, the postulated scenaria in this work are very conservative as they disregard all barriers which could prevent the dispersion of the Sarin plume into the atmosphere after the explosion.

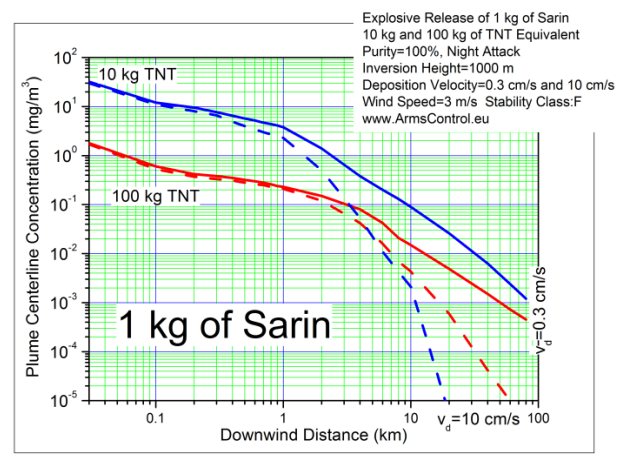

*Figure III.1: Sarin plume centerline concentration with respect to downwind distance from ground zero for a postulated night attack on the alleged chemical weapons site at the Barzah Scientific Research Center. The postulated scenario assumes that 1 kg of Sarin is dispersed by a Tomahawk warhead explosion with two different yields (10 kg TNT equivalent - blue lines) and (100 kg TNT equivalent - red lines) and two different deposition velocities ($v_d$=0.3 m/sec - solid lines) and ($v_d$=10 m/sec - dashed lines). Other meteorological parameters are as follows: Stability Class F, wind speed u=3 m/sec, Inversion Height=1000 m.*

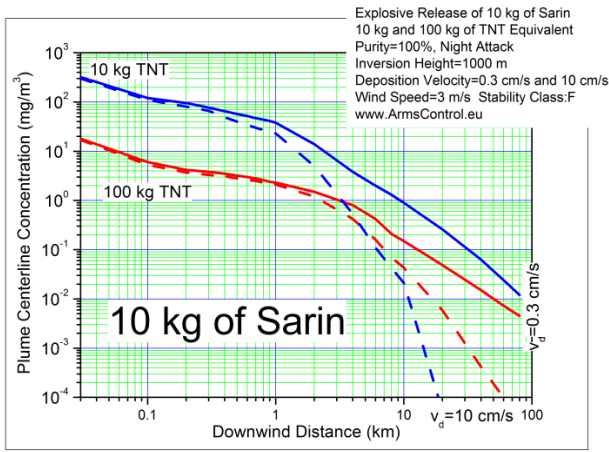

*Figure III.2: Sarin plume centerline concentration with respect to downwind distance from ground zero for a postulated night attack on the alleged chemical weapons site at the Barzah Scientific Research Center. The postulated scenario assumes that 10 kg of Sarin are dispersed by a Tomahawk warhead explosion with two different yields (10 kg TNT equivalent - blue lines) and (100 kg TNT equivalent - red lines) and two different deposition velocities ($v_d$=0.3 m/sec - solid lines) and ($v_d$=10 m/sec - dashed lines). Other meteorological parameters are as follows: Stability Class F, wind speed u=3 m/sec, Inversion Height=1000 m.*

Figures III.1,2,3 show the Sarin centerline concentrations downwind with respect to distance from ground zero for a postulated night attack on the alleged chemical weapons sites in Syria. The postulated scenario assumes that 1 kg, 10 kg and 100 kg of Sarin are dispersed by a Tomahawk warhead explosion of 10 kg TNT equivalent (blue lines) and 100 kg TNT equivalent (red lines) with deposition velocities of $v_d$=0.3 m/sec (solid lines) and $v_d$=10 m/sec (dashed lines) (Stability Class F, wind speed u=3 m/sec, Inversion Height=1000 m are assumed in all cases). Figure III.1 shows that in the event of the explosive dispersion of one kilogram of Sarin with 10 kg of TNT equivalent and a deposition velocity $v_d$=0.3 cm/sec individuals remaining immersed in the Sarin plume at distances up to 4.1 km downwind for ten minutes may experience life threatening effects (AEGL-3 concentration: 0.37 mg/m$^3$).

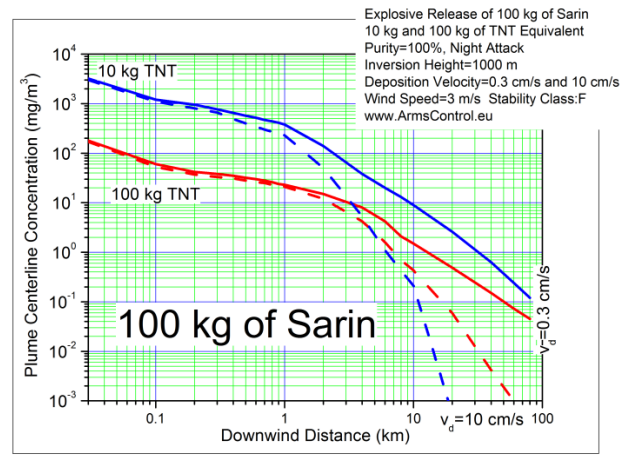

*Figure III.3: Sarin plume centerline concentration with respect to downwind distance from ground zero for a postulated night attack on the alleged chemical weapons site at the Barzah Scientific Research Center. The postulated scenario assumes that 100 kg of Sarin are dispersed by a Tomahawk warhead explosion with two different yields (10 kg TNT equivalent - blue lines) and (100 kg TNT equivalent - red lines) and two different deposition velocities ($v_d$=0.3 m/sec - solid lines) and ($v_d$=10 m/sec - dashed lines). Other meteorological parameters are as follows: Stability Class F, wind speed u=3 m/sec, Inversion Height=1000 m.*

Likewise, under the same input parameters receptors up to a distance of 10 km (AEGL-2 concentration: 0.086 mg/m$^3$) downwind could experience serious health effects (some long-lasting and irreversible) and finally people standing for the same time at distances up to 38 km downwind from ground zero might complain for irritation and notable discomfort (AEGL-1 concentration: 0.0069 mg/m$^3$). If the deposition velocity is larger ($v_d$=10 cm/sec) the above three distances are reduced to 2.2km/3.6km/6.9km respectively (AEGL-3/2/1). According to the same Figure III.1 on the other hand, if one kilogram of Sarin is dispersed with 100 kg of TNT the above AEGL-3/3/1 maximum distances for deposition velocities of 0.3 cm/sec and 10 cm/sec respectively are 0.33 km/3.7 km /16 km and 0.2 km/3 km/8 km. Finally, Figure III.3 suggests that the dispersion of 100 kg of Sarin with 10 kg (100 kg) TNT would extend the AEGL-3/2/1 zones up to distances of ($v_d$=0.3 cm/sec): 50(24) km/94(54) km/191(200+) km and ($v_d$=10 cm/sec): 8.4(10) km/11(17) km/15(36) covering respective areas of ($v_d$=0.3 cm/sec): 26(8.9) km$^2$/78(34) km$^2$/294(303) km$^2$ and ($v_d$=10 cm/sec): 1.6(2.6) km$^2$/3.0(0.027) km$^2$/5.8(0.033) km$^2$.

Naturally, Figure III.1, Figure III.2 and Figure III.3 indicate that the larger the quantities of Sarin released in the postulated scenario the larger the maximum distances and the covered areas for the three AEGL-3/2/1 risk zones. The striking outcome of the previous calculations is that if kilograms (hundreds of kilograms) of Sarin are dispersed from a typical chemical weapons site with small quantities of explosives (~10 kg TNT) and the plume travels uninhibited downwind then there is a probability that unprotected people who remain immersed in the cloud downwind for at least ten minutes may experience lethal or serious adverse health effects tens (hundreds) of kilometers away from ground zero. Owing to the large errors associated with the direction of the wind, risk assessment and predictions as well as evacuation and mitigation decisions should not be based exclusively on the idealized cigar-shaped isodose contours of the Gaussian model. Rather, to eliminate uncertainties and errors associated with wind direction variability a 360° potential hazard circular zone should also be mapped and considered. Accordingly, Figure III.4. shows three concentric circular risk zones (centered on the Barzah Research Center) AEGL-3 (red), AEGL-2 (green), AEGL-1 (blue) and consists of two maps adjacent to each other where the lower one is obviously a magnification of the upper one. The three inner circular risk zones (red, green, blue) correspond to the dispersion of 1 kg of Sarin (360° Risk Zone Radii: 4.1km/10km/38km) while the three outer ones (red, green, blue) correspond to the dispersion of 100 kg of Sarin (360° Risk Zone Radii: 104km/153km/200km 50km/94km/191km). Obviously, the radii of the circular risk zones are the maximum distances at which a particular AEGL zone can extend (as predicted by EPIcode). Both scenaria (1 kg and 100 kg of Sarin) assume 10-min exposures, night attacks with an explosion of 10 kg TNT equivalent and the following meteorological parameters: stability class F, wind speed 3 m/sec, Mixing Height 1000 m, deposition velocity 0.3 cm/sec (the lower map only shows the AEGL-3 and AEGL-2 zones).

A more realistic hazard prediction relies on the isodose contours derived by EPIcode which also calculates the areas enclosed by the three AEGL zones for the postulated Sarin explosive releases. Figure III.5 consists of two adjacent Google maps showing the Gaussian model predictions (cigar-shaped isodose contours) for the two postulated Sarin releases from the Barzah Research Center, namely 1 kg of Sarin (above) and 100 kg of Sarin (below). The contours are derived assuming an explosion of 10 kg TNT equivalent and the meteorological conditions that existed during the April 14, 2018 attack (see subsection II.B). Sarin plumes would create three AEGL-3/2/1 zones (10-min exposure) with maximum distances plotted in Figure III.1 and Figure III.3 which would cover areas as follows (1 kg/100 kg): AEGL-3 (0.33km$^2$/26km$^2$) AEGL-2 (1.5km$^2$/78 km$^2$) AEGL-1 (18 km$^2$/294 km$^2$).

The average population density of Syria is about 100 inhabitants per km$^2$, thus to obtain the number of people expected to be inside a particular AEGL risk zone the aforementioned areas should be multiplied by 100. Note that due to the meteorological conditions during the attack the three AEGL zones would probably be covering areas to the north-east of the Barzah Research Center which is a rather sparsely populated area. Fortunately, considering the wind direction during the attack Damascus was located upwind and therefore the capital of Syria would not be likely to experience any significant symptoms if the postulated quantities of Sarin released by the explosion were small (a few kilograms).

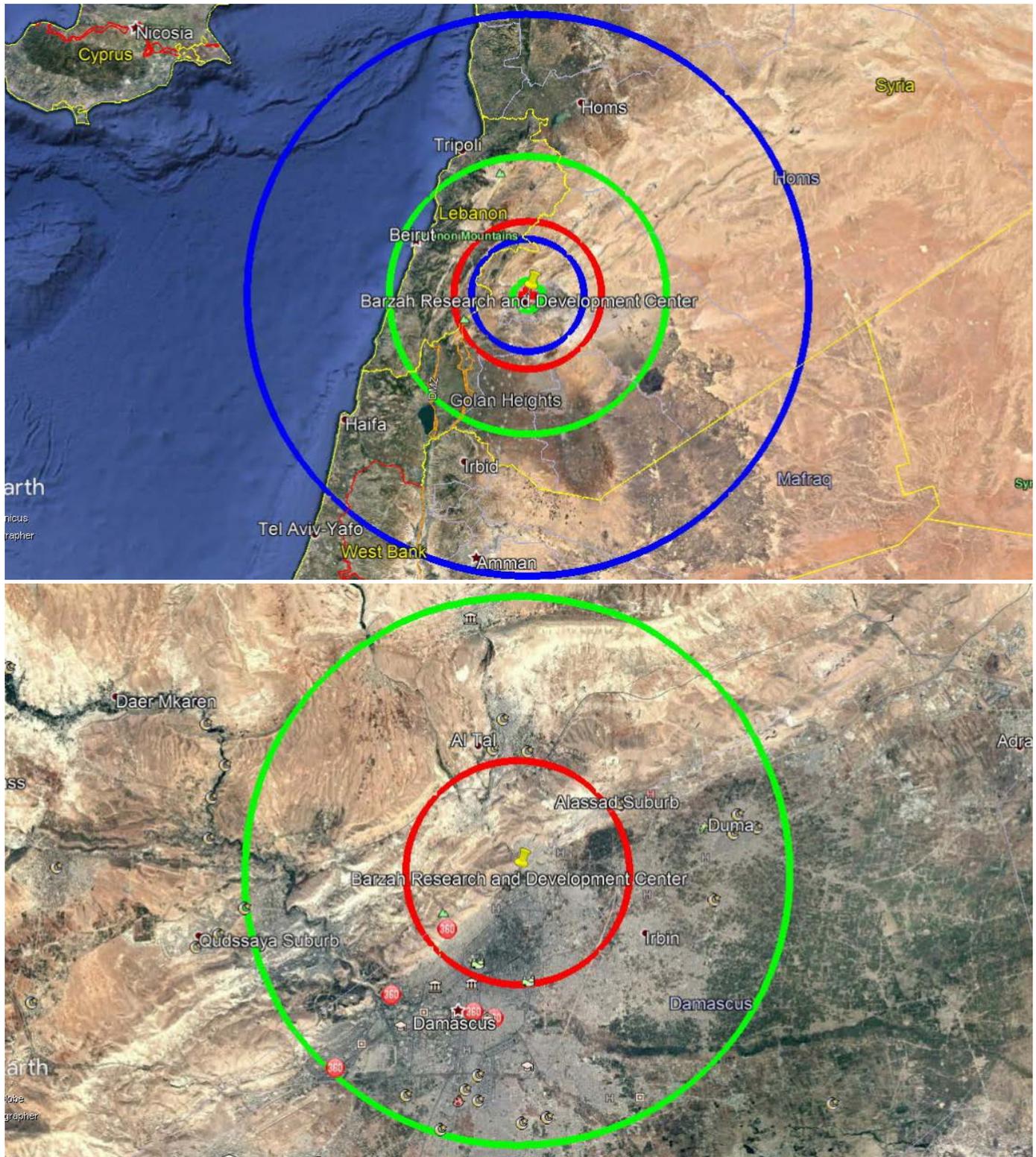

*Figure III.4: Three concentric circular risk zones (centered on the Barzah Research Center) AEGL-3 (red), AEGL-2 (green), AEGL-1 (blue). The three inner circular risk zones correspond to the dispersion of 1 kg of Sarin (Risk Zone Radii: 4.1km/10km/38km) while the three outer ones to the dispersion of 100 kg of Sarin (Risk Zone Radii: 50km/94km/191km). Both Sarin scenaria assume 10-min exposures, night attacks with an explosion of 10 kg TNT equivalent and the following meteorological parameters: stability class F, wind speed 3 m/sec, Mixing Height 1000 m, deposition velocity 0.3 cm/sec (see text). The lower map is a magnification of the area around ground zero and only shows the AEGL-3 and AEGL-2 for the dispersion of 1 kg of Sarin.*

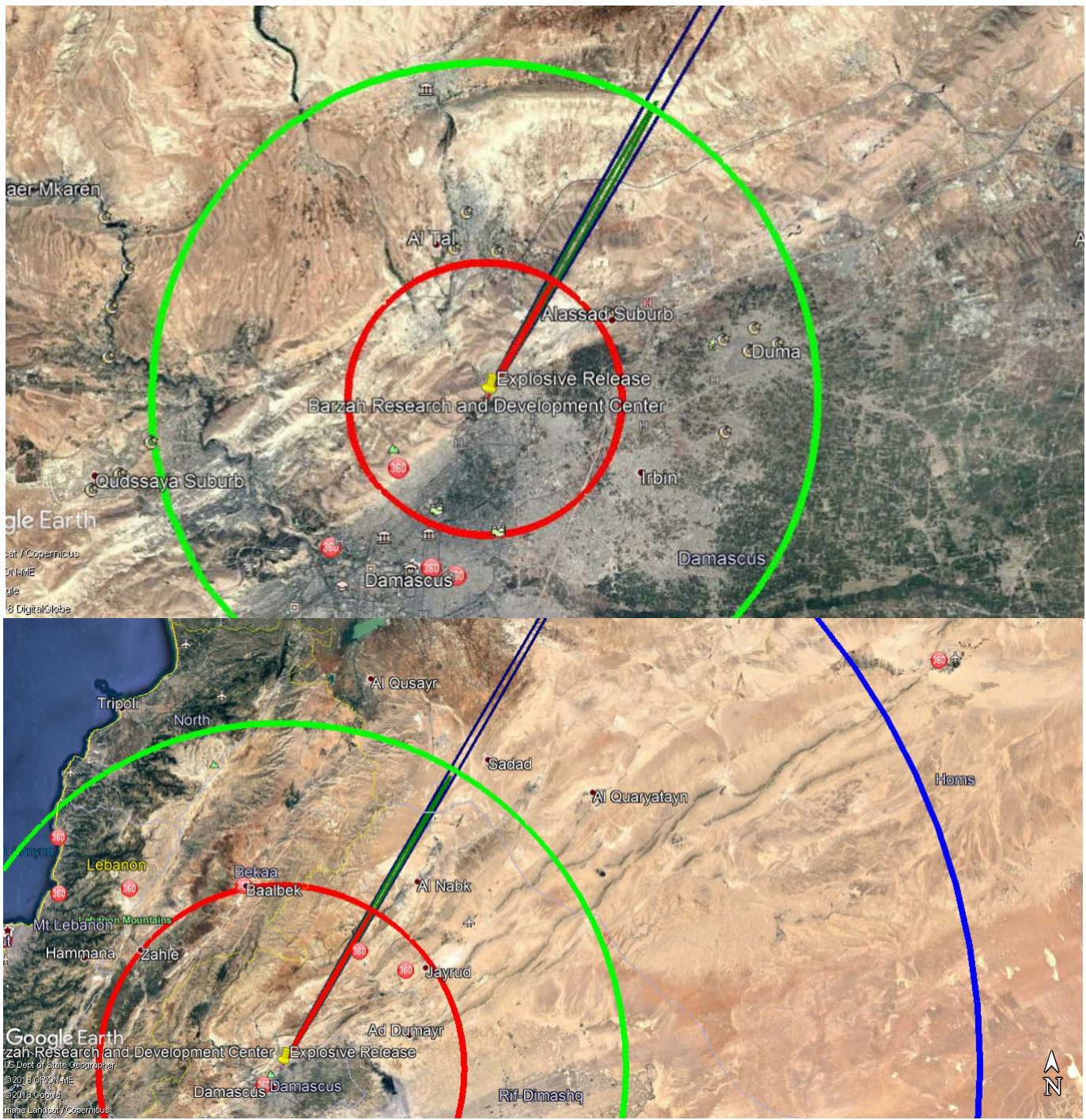

*Figure III.5: EPIcode isodose contours for the explosive release of 1 kg of Sarin (above) and 100 kg of Sarin (below) from the Barzah Research Center assuming an explosion of 10 kg TNT equivalent and the meteorological conditions that existed during the April 14, 2018 attack (see text). Sarin plumes would create three AEGL-3/2/1 zones (10-min exposure) with maximum distances which would cover areas as follows (1 kg/100 kg): AEGL-3 ($0.33 km^2/26 km^2$) AEGL-2 ($1.5 km^2/78 km^2$) AEGL-1 ($18 km^2/294 km^2$). The average population density of Syria is about 100 inhabitants per $km^2$, thus to obtain the number of people expected to be inside a particular AEGL risk zone the aforementioned areas should be multiplied by 100.*

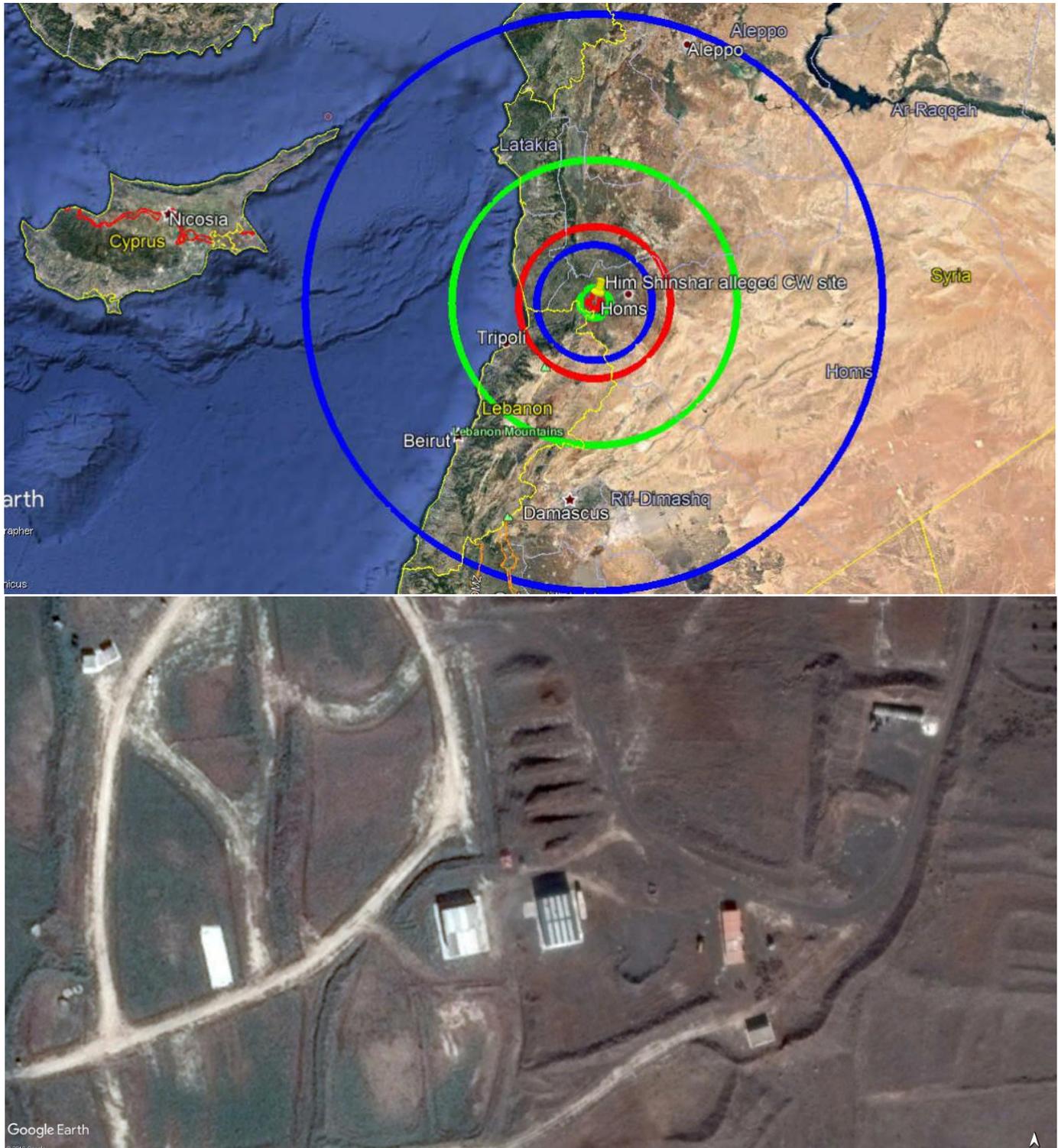

*Figure III.6: The upper map shows three concentric circular risk zones (centered on the Him Shinshar alleged chemical weapons storage facility) AEGL-3 (red), AEGL-2 (green), AEGL-1 (blue). The three inner circular risk zones correspond to the dispersion of 1 kg of Sarin (Risk Zone Radii: 4.1km/10km/38km) while the three outer ones to the dispersion of 100 kg of Sarin (Risk Zone Radii: 50km/94km/191km). Both Sarin scenaria assume 10-min exposures, night attacks with an explosion of 10 kg TNT equivalent and the following meteorological parameters: stability class F, wind speed 3 m/sec, Mixing Height 1000 m, deposition velocity 0.3 cm/sec (see text). The lower map is a close-up image of the targets before the attack.*

However, even if only a few kilograms of Sarin had been explosively released then people living a few kilometers to the north-east of the Barzah research center would most likely experience at least some notable discomfort.

Regarding the attack on the alleged chemical weapons storage facility near Homs (Him Shinshar) an implementation of the EPIcode models yields the concentric circular risk zones AEGL-3 (red), AEGL-2 (green), AEGL-1 (blue) which are now centered on the Him Shinshar installation (see Figure III.6). Note that under certain weather conditions the explosive dispersion of one kilogram of Sarin from Him Shinshar could affect people as far as Tartus while under the recorded weather conditions even such a small amount of Sarin could cause irreversible adverse health effects on people living in Homs. Likewise, the explosive dispersion of 100 kg of Sarin from Him Shinshar could cause serious adverse health effects in Damascus, which however at the time of attack was located upwind which minimized the risk of exposure.

## IV. CONCLUSIONS (EXECUTIVE SUMMARY)

As this research report is expected to be of interest to a large audience with limited background on chemical weapons effects it is imperative that the conclusions should also be in the form of an executive summary. The criminal use of chemical weapons in Syria is undeniable; however the perpetrators have not been identified beyond reasonable doubt. In this study a series of simulations are performed to predict and assess the effects of explosive releases of Sarin from chemical weapons sites such as research centers, factories and ammunition depots. The simulations are carried out by means of the EPIcode software while its input parameters governing the postulated attacks and/or accidents on a chemical weapons site are constrained within realistic ranges of values. The most critical constraints are the time of attack, the relevant meteorological conditions during the attack, the yield of the explosion and the source term (quantity) of Sarin dispersed. Regarding large Sarin storage sites, according to the simulations analyzed in detail in this work the explosive release of 100 kg of Sarin with 100 kg of TNT (e.g. a Tomahawk missile explosion) during a night attack with a gentle breeze and a moderately stable atmosphere could cause life-threatening health effects or death (AEGL-3) up to distances of 24 km downwind covering an area of 8.9 km$^2$ and irreversible or other serious, long-lasting adverse health effects (AEGL-2) up to distances of 54 km downwind covering an area of 34 km$^2$ and finally it could cause discomfort, irritation, or certain asymptomatic non-sensory effects (AEGL-1) at distances larger than 200 km downwind covering an area of 303 km$^2$. In the event that 100 tons of Sarin are dispersed with the same quantity of TNT then all the three AEGL-3/2/1 risk zones will exceed a distance of 200 km covering areas as follows AEGL-3 (4000 km$^2$), AEGL-2 (6000 km$^2$), AEGL-1 (8000 km$^2$).

The April 14, 2018 missile attacks carried out by the United States, the United Kingdom and France against alleged chemical weapons sites in Syria alerted the Arms Control Center as these attacks could possibly have catastrophic effects in the area. A great source of concern was that, if the targets were indeed Sarin storage sites, a military attack on them would result in explosive releases of large quantities of Sarin, which is what had happened during the US demolition operations at the Khamisiyah Pit in Iraq (1991) believed to have been a possible source of the "Gulf War Syndrome". Thus, after simulating a random release of 100 kg and 100 tons of Sarin (large storage sites) with a powerful 100 kg TNT explosion (compatible with yields expected in military and terrorist attacks) this work focused on a case study, namely the April 14, 2018 missile attacks against Syria. Using appropriate meteorological and weapons data all the parameters of the attacks were restricted within a reasonable range of values. For each attack a reasonable explosive yield range was postulated which could result from a combination of missile and chemical munitions explosions (terrorist attacks and accidents are obviously covered by the same scenaria). The simulations assumed that the night attacks on the alleged

Sarin weapons sites in Syria (Barzah Research Center, Him Shinshar) resulted in the explosive dispersion of 1 kg, 10 kg and 100 kg of Sarin with 10 kg TNT and 100 kg TNT, respectively, adopting the meteorological parameters that existed at the time of attack and two reasonably constraining values for the deposition velocity (0.3 cm/sec, 10 cm/sec). Sarin plume centerline concentrations were plotted with respect to distance downwind for the above parameters and the results were also illustrated on Google Earth Maps in two forms: Three AEGL-3/2/1 concentric circular risk zones around ground zero and two AEGL-3/2 isodose contours, with the former showing a 360° risk zone to allow for wind direction uncertainties and the latter showing a more precise footprint of the AEGL zones. In the main text there are detailed data and analysis for each particular simulation of this report, however the results mapped in Figure III.4, Figure III.5, and Figure III.6, illustrate vividly the dimensions of the postulated hazard which, given the uncertainties of the Gaussian models, can be expressed as follows: If kilograms (hundreds of kilograms) of Sarin are dispersed from a typical chemical weapons site with an explosion of few kilograms TNT and the plume travels uninhibited downwind then there is a probability that unprotected people who remain immersed in the cloud downwind for at least ten minutes may experience lethal or serious adverse health effects tens (hundreds) of kilometers away from ground zero and the corresponding risk zones can cover a few square kilometers (a few hundred square kilometers).

For example the explosive dispersion of (1 kg/100 kg) Sarin with 10 kg of TNT would create three AEGL-3/2/1 zones which would cover areas as follows: AEGL-3 (0.33km$^2$/26km$^2$) AEGL-2 (1.5km$^2$/78 km$^2$) AEGL-1 (18 km$^2$/294 km$^2$) and would extend up to distances of AEGL-3 (4.1km/50km), AEGL-2 (10km/94km) and AEGL-1 (38km/191km). The average population density of Syria is about 100 inhabitants per km$^2$, thus to obtain the number of people expected to be inside a particular AEGL risk zone the aforementioned areas should be multiplied by 100.

The undeniable scientific result of this report is that even if a few kilograms of Sarin had been explosively released from the alleged chemical weapons sites targeted in Syria then hundreds to thousands of people would have experienced lethal or serious irreversible health effects. Moreover, if the Sarin released at the Khamisiyah Pit in Iraq (1991) is indeed a source of the "Gulf War Syndrome" then the April 14, 2018 attacks against the alleged Sarin sites in Syria might have generated a similar "Syrian War Syndrome". The international community should keep monitoring the targeted areas in Syria for possible symptoms. However, regardless of the release of Sarin from the targeted sites it is obvious that the repeated criminal use of Sarin in Syria may give rise to a "Syrian War Syndrome" anyway, possibly aggravated by the April 14, 2018 attacks.

APPENDIX

There is no appendix to this study

ACKNOWLEDGMENT

The authors acknowledge useful discussions with their colleagues at the Hellenic Military Academy and the Arms Control Center: Assist. Professor Dr. Paraskevi Divari, Assist. Professor Dr. Georgia Melagraki and Dr. Aggelos Vorvolakos (all members of the Editorial Board of the Arms Control Journal)

CONFLICT OF INTEREST STATEMENT

The authors of this work certify that they have no affiliations with or involvement in any organization or entity with any financial interest or non-financial interest (e.g. political, religious etc.) in the subject matter or materials discussed in this manuscript.

BIOGRAPHIES

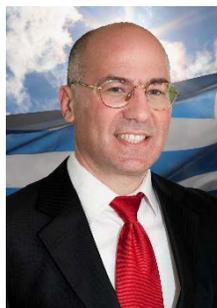

Professor Dr. Theodore Liolios is the Founder and Director of the Arms Control Center. He was awarded a BSc in physics and a PhD in nuclear and atomic physics from the University of Thessaloniki (Greece). After receiving his PhD he was awarded various post-doc fellowships and scholarships including the extremely competitive Marie Curie fellowship by the European Commission in the field of Nuclear Astrophysics (Aarhus University, Denmark). As a young lecturer while on a sabbatical leave he spent a semester at the CISSM (University of Maryland) invited by Professor Dr. Steve Fetter where he worked on nuclear weapons accidents effects (focusing on the 2007 USAF "Bent Spear" Nuclear Weapons Incident). He is now full Professor of Nuclear Physics and Military Sciences in the Department of Military Sciences at the Hellenic Military Academy and is also the Director of the Physical Sciences Department at the Hellenic Military Academy. He began his career in military education when he was appointed by the Hellenic Army General Staff adjunct professor at the Special Weapons Chair of the Hellenic Supreme War College (in 1998 at the age of 28 soon after receiving his PhD in nuclear physics). He has written two books in arms control and weapons effects (in Greek, one received the Hellenic Navy Research Prize) which have favorable reviews by the Chief of the Hellenic Army General Staff and the Hellenic National Defense General staff and various renowned scientists and dignitaries). He has numerous military science articles (in Greek, published on the website of the Arms Control Center) and various published papers in peer-reviewed scientific journals (18 of them are single-authored) in nuclear physics, atomic physics and nuclear astrophysics (Phys.Rev.C, Nucl.Phys.A, J.Phys.A etc.). His teaching and research are focused on (a) nuclear-atomic physics, (b) homeland & international security, (c) counter-terrorism, (c) special weapons effects, sciences and defenses, and (d) arms control and non-proliferation. He has taught weapons science courses to thousands of Hellenic Army military officers for two decades while lots of his former students have become generals (Chiefs and Deputy Chiefs of the Hellenic Army General Staff., etc). He has supervised many graduation theses (both at an undergraduate and a postgraduate level) in higher military educational institutions and the merchant marine academy (where he also served as a full professor of science for a short time at a young age). His excellent research group at the Arms Control Center consists of high quality scientists (tenured or adjunct members of faculty at the Hellenic Military Academy) and is available for research contracts and collaborations in all aspects of security sciences. He has appeared many times on national and local news channels (tv and radio) in Greece as an invited expert on the above subjects (a-d) (e.g. see sample videos here and here). Professor Dr. Theodore Liolios is available as a visiting or adjunct professor, guest lecturer or conference speaker (online or on-site/on-campus) on the above subjects and especially on the subject matter of the present work (see also a sample of his webinars here).
*(Please visit the website of the Arms Control Center for more information about its Director: www.ArmsControl.eu )*

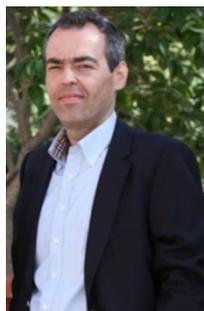

Assistant Professor Dr. Konstantinos Kolovos graduated with honors with a B.Sc and M.Sc. in Chemical Engineering from National Technical University of Athens (Greece), School of Chemical Engineering in 1997 and in 2003 he received his PhD in Chemical Engineering from the same institute, in the field of Cement Chemistry. During his Doctoral Studies he was consecutively receiving scholarships from the Greek State's Scholarship Foundation (1998-2002), Alexander S. Onassis Foundation (2002-2003) and National Technical University of Athens. After receiving his PhD he was awarded a post-doc fellowship and scholarship from the Greek State's Scholarship Foundation (2005-2006). He has participated in various Research Programs in the field of cement and concrete chemistry and technology, in several international congresses concerning cement chemistry and is the author of 35 research papers published in international peer-reviewed scientific journals, several of them awarded by the Greek Thomaidion Award. He is currently an Assistant Professor of Chemistry at Hellenic Army Academy and a Senior Researcher at the Arms Control Center.

DISCLAIMER